\documentclass[preprint,showpacs,preprintnumbers,amsmath,amssymb,
groupedaddress,floatfix,tightenlines]{revtex4}
\usepackage{graphicx}
\usepackage{color}

\begin{document}


\title{Electroweak symmetry breaking and cold dark matter \\
from strongly interacting hidden sector}



\author{Taeil Hur}
\affiliation{Department of Physics, KAIST, Daejon 305--701, Republic of Korea}

\author{Dong-Won Jung}
\email[]{dwjung@phys.nthu.edu.tw}
\affiliation{Physics Department and CMTP, National Central University, Jhongli, , 32054, Taiwan }
\affiliation{Department of Physics, National Tsing Hua University, Hsinchu 300, Taiwan
}

\author{P. Ko}
\email[]{pko@kias.re.kr}
\affiliation{School of Physics, KIAS, Seoul 130--722,Republic of Korea}

\author{Jae Yong Lee}
\affiliation{Department of Physics, Korea University, 
Seoul 136--701, Republic of Korea}


\date{\today}

\begin{abstract}
We consider a hidden sector with a vectorlike confining gauge 
theory like QCD with $N_{h,c}$ colors and $N_{h,f}$ 
light quarks ${\cal Q}_h$ in the hidden sector. 
Then a scale $\Lambda_H$ would be generated by dimensional transmutation,  and chiral symmetry breaking occurs in the 
hidden sector. 
This scale $\Lambda_H$ can play a role of the SM Higgs mass parameter,  triggering electroweak symmetry breaking (EWSB). 
Furthermore the lightest mesons in the hidden sector is stable 
by flavor conservation of the hidden sector strong interaction, 
and could be a good cold dark matter (CDM). We study collider phenomenology, and relic density and direct detection rates 
of the CDM of this model. 
\end{abstract}

\pacs{}
\keywords{QCD, chiral symmetry breaking, cold dark matter, electroweak symmetry breaking, hidden sector}

\maketitle



Understanding the origin of the electroweak symmetry breaking (EWSB)  is
one of the key issues in particle physics.  Experimental searches 
for it have been made at LEP and Fermilab Tevatron, and are being planned at LHC.  We would like to understand why 
$v \ll M_{\rm Pl}$ at tree level,  as well as how to avoid fine 
tuning of the quadratic divergence to the Higgs (mass)$^2$ parameter at each order of perturbation theory.
The latter problem have been a major motivation for many models beyond
the standard model (SM). On the other hand the former problem did not receive much attention
within the SM. It would be nice if we can understand why the electroweak scale is
so small compared to the Planck scale in a similar manner as we understand
why proton mass is so small compared with the Planck scale in QCD.
Quantum fluctuation of quark and gluon fields makes proton massive, and 
naturally lighter than Planck scale within QCD: $m_p \sim O(\Lambda_{\rm QCD})
\ll M_{\rm Pl}$.
It is a tantalizing question if one can achieve the same for EWSB scale 
\cite{Hill:2005wg,Wilczek:2005ez}.
Technicolor (TC) provides an answer for this question, but its  naive
version is in serious conflict with the electroweak precision test (EWPT) \cite{stu}.

Another important problem in particle physics and cosmology is to
reveal the nature of nonbaryonic cold dark matter (CDM), 
which makes $\sim 25 \%$ of the total energy density of the universe.  Nonbaryonic CDM clearly calls for physics beyond the 
SM,  since there are no good candidates for it within the SM.  
There are many particle physics models for CDM, among which the lightest supersymmetric particle (LSP) such as neutralino or 
gravitino is presently the most popular candidate. 
In many cases, the stability of a CDM is put in by hand by imposing some $Z_2$ symmetries: {\it e.g.}, $R-$parity in 
supersymmetry (SUSY) models. 
It would be nice if the CDM is stable by an accidental symmetry of 
the underlying local gauge symmetry, like baryon/lepton number 
conservations in the SM, and not by an ad hoc $Z_2$ symmetry. 
Within the SM, one can write down all the possible terms that are
consistent with global Poincar${\rm \acute{e}}$ and local gauge symmetries with
renormalizability. Then the baryon/lepton number conservations follow as accidental symmetries. Another example of an 
accidental symmetry is the isospin symmetry in QCD due to 
$(m_u - m_d) \ll \Lambda_{\rm QCD}$.

One can ask if it would be possible to generate EWSB and its scale 
as one generates proton mass in QCD 
\cite{Hill:2005wg,Wilczek:2005ez}, and also accommodate a 
stable CDM candidate in the same framework.
We argue that the answer to this question is affirmative, 
if there is a new vectorlike confining strong interaction with 
$N_{h,c}$ colors and $N_{h,f}$ flavors in the hidden sector,
\begin{equation}
{\cal L}_{\rm hidden} = - {1\over 4} {\cal G}_{\mu\nu} {\cal G}^{\mu\nu}
+ \sum_{k=1}^{N_{h,f}} \overline{\cal Q}_k \left( i D\cdot \gamma
- M_{{\cal Q}_k} \right) {\cal Q}_k .
\end{equation}
The hidden sector quarks and gluons are SM singlets, and have no 
renormalizable interactions with the SM particles. Interactions with the SM sector would occur through exchanges of some messengers,
which would be model dependent.  Such effects would be described 
by  higher dimensional operators below the messenger scales, 
and are suppressed by powers of $1/\Lambda_{\rm mess}$, 
where $\Lambda_{\rm mess}$ is a messenger scale, at which  interactions between the SM and the hidden sectors are generated. 
In this letter, we don't specify the messenger in detail, relegating 
the discussions to the future works \cite{progress}.

Since the hidden sector gauge theory is a strongly interacting 
theory, we have to construct an appropriate effective field theory for
given numbers of colors and flavors in the hidden sector. 
The low energy effective theory of this hidden sector gauge theory will be something similar to (non)linear sigma models of QCD. 

There are two key points of this letter. First of all, a dynamical scale 
$\Lambda_H$ (naturally smaller than $M_{\rm Pl}$ like in QCD) is generated quantum mechanically in the hidden sector by dimensional transmutation, 
and spontaneous chiral symmetry breaking occurs.  This scale 
$\Lambda_H$ can play a role of Higgs mass parameter in the SM, 
if appropriate conditions are met.
Chiral symmetry breaking in the hidden sector is the same as the usual technicolor models, except that the hidden sector quarks 
are SM singlets,  rather than the SM doublets.  Therefore the EWPT constraints become milder in our scenario \cite{progress}. 
We can keep nicety of technicolor idea in a new form.    
Secondly, the lightest mesons and baryons in the hidden sector 
can be good candidates for CDM, and their stability is guaranteed 
by  accidental chiral symmetry or hidden sector baryon number conservation, rather than by ad hoc $Z_2$ symmetry. 
This is a clear advantage of our model for CDM 
compared with many other CDM candidates.

Physics with a hidden sector is by no means a new idea.
There are numerous works on physics with a hidden sector
\cite{Ignatiev:2000yy,Berezhiani:2003xm,Barbieri:2005ri,Strassler:2006im,
Patt:2006fw}, and also in SUSY and superstring theory.
A new ingredient in this letter compared with earlier literature is
that we consider technicolor in the hidden sector, the weak scale of the SM sector is generated in the hidden sector by dimensional transmutation,  and the CDM is the lightest hadrons  
in the hidden sector and is composite.

Our picture is similar to SUSY models, where SUSY is broken spontaneously by hidden sector gaugino condensation. 
Then SUSY breaking effect is transmitted to the SM sector by messengers. Likewise in our model,  a scale is generated 
in the hidden sector by dimensional transmutation (and chiral symmetry breaking), and its effect is transmitted 
to the SM sector by messengers. 


The SM Lagrangian is given by the usual terms:
\begin{widetext}
\begin{equation}
  \label{eq:sm}
  {\cal L}_{\rm SM} 
= {\cal L}_{\rm kin} -
{\lambda_1 \over 2}~( H_1^{\dagger} H_1 )^2
+ {\mu_1^2}~ H_1^{\dagger} H_1
- \left( \overline{Q}^i H_1 Y^D_{ij} D^j + 
\overline{Q}^i \tilde{H}_1 Y_{ij}^U U^j
+ \overline{L}^i H_1 Y^E_{ij} E^j +h.c.  \right)
\end{equation}
\end{widetext}
where ${\cal L}_{\rm kin}$ is the kinetic term for the SM chiral fermion, Higgs boson and three gauge boson fields. We have suppressed the neutrino mass terms since they are irrelevant 
to our main points. In the following, we consider Eq.s~(1) and (2) 
and discuss EWSB and DM from the strongly interacting hidden sector. 


Far below the hidden sector chiral symmetry breaking scale 
$\Lambda_{h,\chi} \approx 4 \pi \Lambda_H$, the effective 
field theory of the hidden sector strong interaction would be
(non)linear sigma models, similarly to the ordinary QCD. 
The detailed spectra depend on $N_{h,c}$ and $N_{h,f}$ 
with approximate $SU(N_{h,f})_L \times SU(N_{h,f})_R$ chiral
symmetry. One can choose some $N_{h,f}$ and study the phenomenology, and we assume that $N_{h,f}$ is small enough 
so that there is chiral symmetry breaking and confinement 
in the hidden sector.

For illustration purpose, let us consider $N_{h,f}=2$ with 
small current quark masses 
$M_{{\cal Q}_{i=1,2}} \ll \Lambda_H$, and $N_{h,c} = 3$, 
like ordinary QCD with two light flavors. 
In this case, the low energy effective theory is the 
Gell-Mann-Levy(GML)'s linear $\sigma$ model, or nonlinear 
$\sigma$ model with approximate global 
$SU(2)_L \times SU(2)_R$ symmetry.
The relevant degrees of freedom would be the hidden sector pions 
($\pi_h$) and it scalar partner $\sigma_h$,
which forms $SU(2)_L \times SU(2)_R$ bidoublet $H_2$, transforming as 
\[
H_2 (x) \equiv \left( \sigma_h (x) + i \pi_h^a (x) \tau^a \right) 
\rightarrow L_h H_2 (x) R_h^\dagger
\]
under the hidden sector global $SU(2)_L \times SU(2)_R$ 
symmetry.   Here $\tau^a$'s are the Pauli matrices. 
There are also the hidden sector nucleons $N_h = ( p_h , n_h )$, 
which is rather heavy $M_{N_h} \sim N_{h,c} \Lambda_H$. 
In this letter, we ignore the hidden sector nucleons for simplicity, and consider physics related with $\pi_h$'s and $\sigma_h$, 
since we do not have good theoretical tools to calculate the 
$N_h \overline{N_h} \rightarrow$ (SM particles), which is 
relevant to thermal relic density of the hidden sector baryons.
For the case of hidden sector pions which, the low energy effective theory of the hidden sector below the hidden sector chiral symmetry breaking scale $\Lambda_{h,\chi} \sim 
4 \pi v_2$ (where $v_2 \equiv \langle \sigma_h \rangle \sim \Lambda_H$) 
would be the Gell-Mann-Levy's linear $\sigma$ model, 
or its nonlinear version if $m_{\sigma_h} \gg v_2$. 
For the sake of simplicity, we assume that the low energy effective 
theory of the hidden sector is just renormalizable GML 
$\sigma$ model defined in terms of a scalar doublet $H_2$,
postponing the discussions of the higher dimensional operators and 
the nonlinear sigma model version in the future publication
\cite{progress}.  

Then, the potential for $H_1$ and $H_2$ is given by 
\begin{eqnarray}
V(H_1,H_2) = -\mu^2_1 (H^\dagger_1 H_1)+\frac{\lambda_1}{2} 
(H^\dagger_1 H_1)^2
-\mu^2_2 (H^\dagger_2 H_2) \nonumber  \\
 + \frac{\lambda_2}{2} (H^\dagger_2 H_2)^2
+\lambda_3 (H^\dagger_1 H_1)(H^\dagger_2 H_2) + \frac{a v^3_2}{2}\sigma_h 
\end{eqnarray}
where $\lambda_{1,2}>0$ and $\lambda_1 + \lambda_2 + 2 \lambda_3 > 0$.  This potential looks exactly the same as the
ordinary two-Higgs doublet model, except for the $a$ term 
that generates nonzero masses for the hidden sector pions.

Since $H_2$ does not have renormalizable couplings to the SM fermions, 
there is no Higgs-mediated FCNC problem. 
Due to the chiral symmetry breaking in the hidden sector,
$H_2$ develops a VEV $v_2$. Then the $\lambda_3$ term will contribute to the
effective Higgs mass parameter: $m_{H_1}^2 = \mu_1^2 - \lambda_3 v_2^2$.
In particular, one can imagine an interesting possibility that $\mu_1^2 =0$,
and the SM has classical scale symmetry.  Then the entire Higgs mass 
parameter arises quantum mechanically due to dimensional transmutation  
in the hidden sector. In this letter, we concentrate on the scenario
with $\mu_1^2 =0$. General case with $\mu_1^2 \neq 0$ will be discussed 
in Ref.~\cite{progress}.  

The $\lambda_3$ coupling is important in our scenario, 
since it transmits the scale $\Lambda_H$ to the SM Higgs 
sector.  Let us comment on a possible origin of such term.
We can include a real singlet scalar $S$ as a messenger, 
which has the following interactions with the SM Higgs doublet 
and the hidden sector quarks:
\begin{equation}
{\cal L}_{\rm int} = -\frac{1}{2} m_S^2 S^2 + \mu_S S^3 + 
\lambda_S S^2 H_1^\dagger H_1 + 
\sum_{k=1}^{N_{h,f}} \overline{\cal Q}_k \left[ i D\cdot \gamma
- \left( M_{{\cal Q}_k} + \lambda_k S \right) \right] {\cal Q}_k .
\end{equation}
In general, the real singlet scalar $S$ can develop nonzero VEV. 
Integrating out the scalar field $S$,  one can obtain 
\[
{\cal L}_{\rm eff} \sim \langle S \rangle ~
H_1^\dagger H_1 \sum_k \bar{\cal Q}_k {\cal Q}_k 
/ M_S^2 
\]
which in turn becomes the $\lambda_3$ term, after hidden sector chiral symmetry
 breaking. 

Let's look into the interaction more detail. Around the vacuum we
 can write the scalar fields as $ S=\langle S \rangle +\tilde{S}$. 
In the visible sector, 
three vertex $\tilde{S}-H_1-H_1^\dagger$ is generated with coupling 
$\lambda_S\langle S \rangle$ For the hidden sector, we can write down the 
low energy effective Lagrangian which contains the three-vertex $\tilde{S}-
H_2-H_2^\dagger$ with coupling $8 \frac{\Lambda^3_{h,\chi}}{v_2^2}\lambda_k$.
Then the effective vertex $\lambda_3 \left(H_1^\dagger H_1\right) 
\left(H_2^\dagger H_2 \right)$ is derived with 
\begin{equation*}
\lambda_3 = 8 \lambda_S\langle S \rangle 
\frac{\Lambda^3_{h,\chi}}{v_2^2 M_S^2}\lambda_k .
\end{equation*}
More detailed analysis based on the nonlinear realization will be studied in
\cite{progress}.

The Higgs sector of our model is similar to the usual two-Higgs doublet
model, but there are two important differences. 
First of all, the hidden sector Higgs $H_2$ is a singlet under the SM 
gauge group in our model, and EW gauge bosons $W^{\pm}$ and $Z^0$ 
get masses only from the SM Higgs VEV $\langle H_1 \rangle$.
The charged Higgs boson $H^{\pm}$ and the pseudoscalar $A^0$ in the usual
two-Higgs doublet model become dark matter candidates $\pi_h^{\pm,0}$  
in our model, since these particles cannot decay into the SM particles. 
Note that the charge of $\pi_h^{\pm,0}$ is not electric charge, but is 
the 3rd component of the isospin $SU(2)_V$ in the hidden sector.
Secondly, we added the last term $\propto a \sigma_h$ term 
in order to accommodate explicit chiral symmetry breaking in the hidden 
sector from nonzero hidden sector quark masses $M_{{\cal Q}_i}$. 
The resulting hidden sector pions become massive and no cosmological 
problem arises from massless DM in addition to three light neutrinos.

We consider a phase of $V( H_1 , H_2 )$ where both $H_1$ and $H_2$ 
develop the nonzero VEV's:
\begin{equation}
H_1=\left(\begin{array}{c} 0 \\ 
\frac{v_1+h_{\rm SM}}{\sqrt{2}}\end{array}\right),
\qquad
H_2=\left(\begin{array}{c} \pi^+_h \\ 
\frac{v_2+\sigma_h +i \pi_h^0}{\sqrt{2}}
\end{array}\right).
\end{equation}
In order to achieve a correct EWSB in the SM sector, 
we have to satisfy an inequality $\lambda_1 (\lambda_2 +a/2 ) \equiv 
\lambda_1 \lambda_2^\prime  > \lambda^2_3$.

The mass matrix of $ h_{\rm SM}$ and $\sigma_h$ is given by 
\begin{equation}
{\cal M}^2 =  
\left(\begin{array}{cc} \lambda_1 v^2_1 & \lambda_3 v_1v_2 \\
\lambda_3 v_1v_2 & \lambda^\prime_2 v^2_2\end{array}\right)
\end{equation} 
in the $( h_{\rm SM} , \sigma_h )$ basis. 
The mass eigenstates $(h , H)$ are linear combination of $h_{\rm SM}$ and 
$\sigma_h$ that diagonalizes the above mass matrix:
\begin{equation}
\left(\begin{array}{c} h \\ H \end{array}\right)=
\left(\begin{array}{cc}\cos\alpha & \sin\alpha \\ 
-\sin\alpha &\cos\alpha\end{array}\right)
\left(\begin{array}{c} h_{\rm SM} \\ \sigma_h \end{array}\right).
\end{equation}


There are 4 independent parameters, $\tan\beta$, $m_{\pi_h}$,
$\lambda_1$ and $\lambda_2$, and we can trade the latter two by  
$m_{h}$ and $m_{H}$ with $m_h \leq m_H$. In order to demonstrate the main 
points of this letter, we fix $\tan\beta =1$ ($v_1 = v_2$) and scan 
over other parameters with $m_{\pi_h} \lesssim 0.5 \Lambda_H \sim 5 v_2$, 
so that the hidden sector pions can be still regarded as pseudo Goldstone 
bosons. For $\lambda_1$ and $\lambda_2$, we scan upto $\sim 4 \pi$, 
and check if the perturbative unitarity for $W_L W_L$ scattering is
satisfied or not.
The most important constraint on our model comes from the Higgs
boson search and the relic density of the DM, which are calculated
by modifying the micromegas \cite{micromegas}. 

\begin{figure}
\includegraphics[width=10cm] {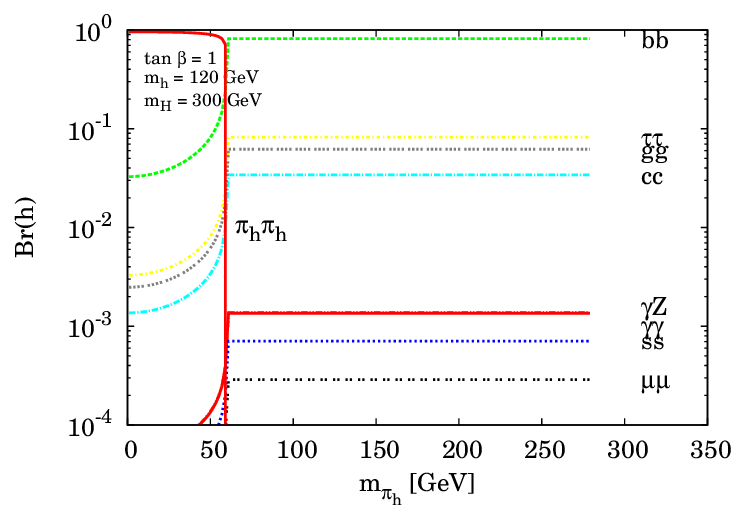}
\includegraphics[width=10cm] {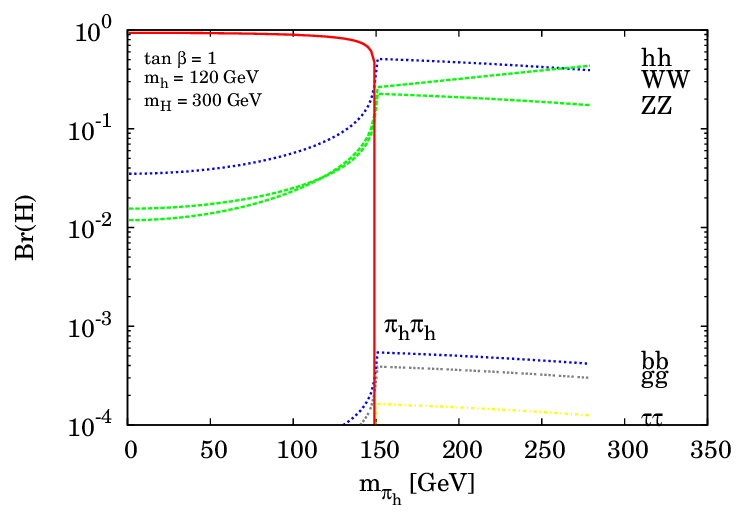} 
\caption{\label{fig:br-t1}
Branching ratios of (top) $h$ and (bottom) $H$ as functions of $m_{\pi_h}$ for 
$\tan\beta = 1$, $m_h = 120$ GeV and $m_H = 300$ GeV.} 
\end{figure}

In the low energy, there are two scalars $h_{\rm SM}$ and $\sigma_h$ which 
mix through $\lambda_3$ term, and make two Higgs bosons $h_1$ and $h_2$.
Note that only $h_{\rm SM}$ couples to the SM gauge bosons and SM fermions.
Therefore the couplings of $h$ and $H$ are modified by $\cos\alpha$
and $\sin\alpha$ of the SM couplings, and it is straightforward to calculate
their decay rates and branching ratios with a few notable differences from
the SM.
Invisible decay channels of neutral scalars $h, H$ can open up through 
$h, H \rightarrow \pi_h \pi_h$, whose 
decay rates depend on the $\pi_h$ masses and other parameters.
This will make more difficult to observe the Higgs bosons $h$ or
$H$ using the visible final states, such as $b \bar{b}$ or 
$t \bar{t}$, and $W^{(*)} W$ or $Z^{(*)} Z$. 
If $m_H > 2 m_h$, then a new channel $H \rightarrow h h$ 
can open up, and the decay width of $H$ will increase. 
In Fig.~\ref{fig:br-t1}, we show the branching ratios for two body 
decays of $h$ of 120 GeV and $H$ with 300 GeV for $\tan\beta = 1$.  
Note that the invisible branching ratios of $h$ and $H$ into 
$\pi_h \pi_h$ could be substantial if $\pi_h$ is light. 

The production cross sections of $h$ and $H$ at Tevatron, LHC or ILC 
are the same as the SM Higgs boson production rates, except that 
the rates are scaled by the $h(H)-t-t$ or $h(H)-V-V$ (with $V=W,Z$)
couplings, which are $\cos^2 \alpha$ and $\sin^2 \alpha$, respectively. 
Therefore the production rates of $h$ and $H$ are  always suppressed 
relative to the SM Higgs production of the same mass.

In Fig.~2, we show the relic densities of $\pi_h$ in the 
$( m_h , m_H )$ plane for $\tan\beta = 1$ with different colors for 
$\log_{10} (\Omega_{\pi_h} h^2)$ between $-6$ and $8$.  
We imposed only $\Omega_{\pi_h} h^2 < 0.122$  (the WMAP bound), since there 
could be additional DM's, namely the lightest hidden sector nucleons or 
axions. The white region is excluded by direct search limit on Higgs boson 
and the correct EWSB vacuum for the SM sector. In a certain parameter 
space, the relic density is too large. This would be cured if we include 
nonrenormalizable interactions between the SM and the hidden sector
\cite{progress}: for example, 
$
\bar{f} f H_2^\dagger H_2$, and other operators. However we may lose  
predictability because of new couplings for these operators.

\begin{figure}
\includegraphics[width=10cm] {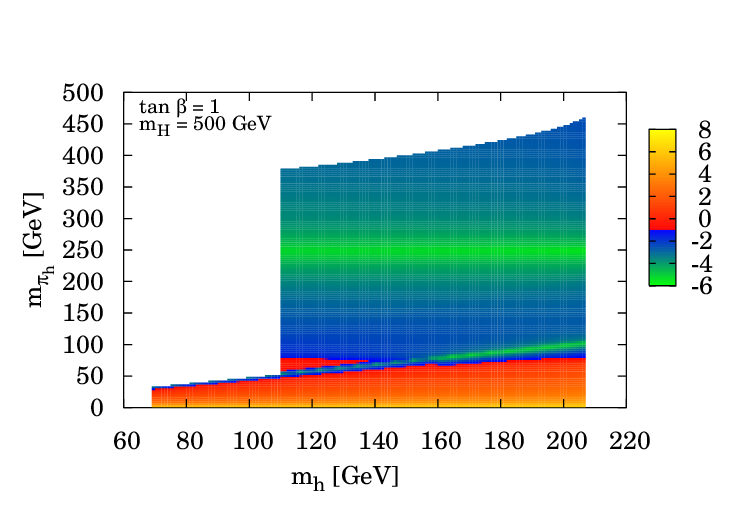}
\caption{
$\Omega_{\pi_h} h^2 $ in the $( m_{ h_1 } , m_{\pi_h} )$ plane  
for $\tan\beta = 1$ and $m_H = 500$ GeV.} 
\end{figure}


In Fig.~3, we show the spin-independent dark matter scattering cross 
section with proton $\sigma_{\rm SI}$ for $\tan\beta = 1$, with the
current bounds from CDMS-II and XENON as well as the projected
sensitivities of XMASS and SuperCDMS. Blue and green dots
denote the relic density $0.096 < \Omega_{\rm DM} h^2 < 0.122$ and 
$\Omega_{\rm DM} < 0.122$. 
Note that our model predicts $\sigma_{\rm SI}$ in a very interesting range,
partially excluded by the current CDMS-II and XENON data, and also
could be tested by the near future experiments. 

\begin{figure}
\includegraphics[width=10cm] {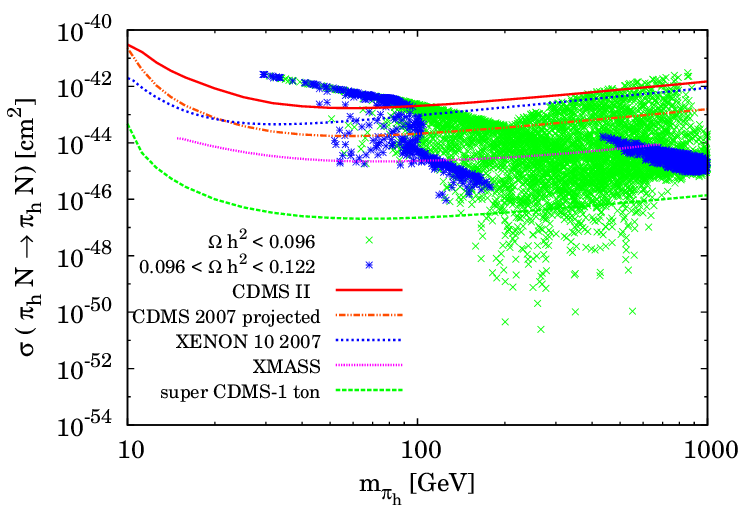}
\caption{\label{fig2}
$\sigma_{SI} (\pi_h p \rightarrow \pi_h p )$ as functions of $m_{\pi_h}$ 
for $\tan\beta = 1$. (For interpretation of the
references to color in this figure, the reader is referred to the web version of this
Letter.)} 
\end{figure}


In summary, we proposed a new possibility that there is a
hidden sector with a new vectorlike confining strong interaction
like QCD.  
A scale $\Lambda_{H}$ is generated quantum mechanically through 
dimensional transmutation,
and it is naturally smaller than the Planck scale. This scale is
transmitted to the SM sector Higgs mass parameter through some messengers.
The lightest hadrons in the hidden sector, $\pi_h^{\pm,0}$, can make
good candidates for the CDM. Stability of CDM in our model is guaranteed 
by flavor symmetry which is an accidental symmetry of the hidden sector 
gauge theory, and not by ad hoc $Z_2$ symmetry in many other models for DM's. 
Direct rate for the DM is around the current or near-future DM 
scattering experiments, and our model predicts detectable rates 
in a sizable parameter space. 
There are two Higgs bosons, one ($h_{\rm SM}$) fundamental in the SM 
sector and the other ($\sigma_h$) composite from the hidden sector. 
These two mix with each other, making two mass eigenstates $h$ and $H$, 
and they can decay into the CDM pair. 
Their invisible branching ratio could be substantial depending on the 
masses of Higgs bosons and the DM particle. 

We can extend the current work in various directions. 
One can study cosmological consequences of the lightest hidden sector 
baryons, baryogenesis and the relation between baryon and DM densities.  
Or we can consider different values of $N_{h,f}$. 
Further, if there is an additional weakly interacting gauge symmetry 
in the hidden sector, such as $SU(2)_L$ or $SU(2)_R$, the weak gauge bosons 
will get massive by eating up the hidden sector pions, and could be 
a spin-one CDM. 
If we impose classical scale symmetry in the hidden sector, we need 
additional real singlet scalar $S$ scalar 
that can provide the masses for the hidden sector quarks.  Then 
the whole theory becomes classically scale invariant and all the mass 
scales could be generated quantum mechanically by dimensional transmutation.
More detailed discussions on these topics will be presented elsewhere 
\cite{progress}. 
Still we hope that this letter is enough to convey the underlying ideas 
on the role of hidden sector technicolor interaction 
on the EWSB and the DM physics, and open a new possibility   
in the model building with dynamical EWSB such as techinicolor.

While we were completing this work, we came to know that F. Wilczek
considered the idea of EWSB from hidden sector strong 
interaction with $a=0$ in Eq. (3)~\cite{wilczek}. 
The basic ideas and the qualitative features of the present and 
related models were presented by P. Ko at the 3rd conference on 
``Dark Side of the Universe,'' University of Minnesota, USA 
(June 5--10, 2007) and at KIAS particle physics journal club (July 4, 2007).

\begin{acknowledgments}
P.K. is grateful to M. Drees, C. Hill, D.K. Hong, C.S. Lim,
A. Masiero, C. Munoz, S. Rudaz and F. Wilczek for useful discussions.
P.K. is supported in part by National Research Foundation
(NRF) through Korea Neutrino Research Center (KNRC) at Seoul National
University. J.Y. Lee was supported by Basic Science Research
Program through NRF grant funded by the MEST (2010-0012779),
and by Mid-career Researcher Program through NRF grant funded
by the MEST (2010-0027811).
\end{acknowledgments}

\end{document}